\documentstyle[psfig]{mn2e}
\newif\ifAMStwofonts
\newcommand{\lapp}{\mbox{\raisebox{-0.3em}{$\stackrel{\textstyle <}{\sim}$}}}
\newcommand{\gapp}{\mbox{\raisebox{-0.3em}{$\stackrel{\textstyle >}{\sim}$}}}
\title[H{\sc i} absorption towards nearby compact radio sources]{H{\sc i} absorption towards nearby compact radio sources}
\author[Yogesh Chandola, S.K. Sirothia and D.J. Saikia]
       {Yogesh Chandola, S.K. Sirothia and D.J. Saikia  \\
National Centre for Radio Astrophysics, TIFR, Pune University Campus, Post Bag 3, Pune 411 007, India \\
}
\date{Accepted.  Received }

\pagerange{\pageref{firstpage}--\pageref{lastpage}}
\pubyear{  }

\begin{document}

\maketitle

\label{firstpage}

\begin{abstract}
We present the results of H{\sc i} absorption measurements towards a sample of 
nearby Compact Steep-Spectrum (CSS) and Giga-Hertz Peaked Spectrum (GPS) radio 
sources, the CORALZ sample, using the Giant Metrewave Radio Telescope (GMRT). 
We observed a sample of 18 sources and find 7 new detections. These sources 
are of lower luminosity than earlier studies of CSS and GPS objects and we 
investigate any dependence of H{\sc i} absorption features on radio luminosity. 
Within the uncertainties, the detection rates and column densities are similar 
to the more luminous objects, with the GPS objects exhibiting a higher detection 
rate than for the CSS objects. The relative velocity of the blueshifted absorption 
features, which may be due to jet-cloud interactions, are within $\sim$$-$250 
km s$^{-1}$ and do not appear to extend to values over 1000 km s$^{-1}$ seen for
the more luminous objects. This could be due to the weaker jets in these objects,
but requires confirmation from observations of a larger sample of sources.
There appears to be no evidence of any dependence of H{\sc i} column density
on either luminosity or redshift, but these new detections are consistent with
the inverse relation between H{\sc i} column density and projected linear size.
\end{abstract}


\section{Introduction}
Radio observations of various classes of active galactic nuclei (AGN) have revealed a wide scale of structures.
For the luminous radio galaxies and quasars these range from the subgalactic-sized Giga-Hertz Peaked Spectrum (GPS)
and Compact Steep-Spectrum (CSS) objects 
(O'Dea 1998) to the giant radio sources which extend over a Mpc in size (e.g. Ishwara-Chandra
\& Saikia 1999).  The GPS sources have convex spectra which peak at $\sim$1 GHz, while the CSS objects may exhibit 
a flattening or turnover at significantly lower frequencies (e.g. O'Dea 1998). The GPS sources have projected 
linear sizes $\lapp$1 kpc, while CSS sources are 
usually defined to have a projected linear size $\lapp$15 kpc (H$_o$ = 71 km s$^{-1}$ Mpc$^{-1}$,
$\Omega_{m}$=0.27, $\Omega_{\Lambda}$=0.73; Spergel et al. 2003). In addition, the high-frequency peakers, 
where the radio spectra peak at frequencies much $>$1 GHz, are likely to exhibit more compact structures 
(e.g. Orienti et al. 2006).
All these sources have steep high-frequency spectra  ($\alpha\gapp$0.5, where $S(\nu)\propto\nu^{-\alpha}$) 
at frequencies beyond their turn-over frequencies. 
On the basis of their somewhat symmetric radio structures, some of these compact sources have also been 
referred to as 
Compact Symmetric Objects or CSOs (e.g. Wilkinson et al. 1994; Taylor et al. 1996; Readhead et al. 1996). It is 
widely believed from both dynamical and spectral ageing studies that the peaked spectrum objects are likely to be 
the youngest with dynamical ages of $\sim$10$^{2}$ to 10$^{3}$ yr (e.g. Phillips \& Mutel 1982; O'Dea 1998; 
Polatidis \& Conway 2003), which evolve into the CSS objects with ages of $\sim$10$^5$ yr 
(e.g. Murgia et al. 1999), and which in turn evolve into the larger sources which could be over 10$^8$ yr old 
(e.g. Jamrozy et al. 2008; Konar et al. 2008).

These young, radio-loud AGN are ideal objects for studying
the triggering of radio activity, the early evolution of classical double-lobed radio sources,
interactions of the radio jets with the interstellar medium, and AGN feedback which could affect
the evolution and formation of galaxies. These sources may also be used as probes of the environments
of the central regions and the interstellar medium of the host galaxies via H{\sc i} 21-cm absorption
observations (e.g. van Gorkom et al. 1989; Vermeulen et al. 2003; Pihlstr\"om, Conway \& Vermeulen 2003;
Gupta et al. 2006, and references therein), as
well as radio polarisation measurements (e.g. Mantovani et al. 1994; Saikia \& Gupta 2003). The absorbing 
neutral hydrogen could be associated with either the circumnuclear disk and the putative torus or the halo
of the host galaxy. Detection of this gas is also important for studying the anisotropy of the radiation 
field and thereby testing the unified scheme for active galaxies. An understanding of the distribution 
and kinematics of this gas over a large range of redshifts and/or luminosities can provide valuable information 
on the evolution of its properties with source size (age), radio luminosity and cosmic epoch.
   
Several recent studies of CSS and GPS objects have demonstrated that H{\sc i} absorption is 
seen in $\sim$35 per cent of the objects, with the H{\sc i} column density being anti-correlated with the 
source size (Vermeulen et al. 2003; Pihlstr\"{o}m, Conway \& Vermeulen 2003; Gupta et al. 2006). To a first 
order, the H{\sc i} gas may be distributed in the form of a disk,
but exhibits a variety of line profiles, suggesting significant and sometimes complex motions. van Gorkom 
et al. (1989) had reported that H{\sc i} absorption tends to be redshifted relative to 
the systemic velocity, suggesting infall
of gas. However, recent observations show the situation to be more complex. These observations find many sources
with substantial blue shifts, suggesting that the atomic gas may also be flowing out, interacting with 
the jets or rotating around 
the nucleus (Vermeulen et al. 2003; Gupta et al. 2006). Morganti et al. (2005) report the detection 
of low optical depth H{\sc i} gas which may be blue shifted by over $\sim$1000 km s$^{-1}$, possibly 
due to jet-cloud interactions.

Gupta et al. (2006) examined and found no evidence of any correlation between H{\sc i} column density 
and luminosity or redshift for a sample of CSS and GPS objects with a radio luminosity largely 
$\geq$10$^{25}$ W Hz$^{-1}$ at 5 GHz. 
Their sample was compiled by combining the results of their observations with existing observations 
in the literature.  Since most of these sources were compiled from strong flux-density limited samples, 
luminosity and redshift were strongly correlated.  To study the properties of H{\sc i} absorbers in 
lower-luminosity GPS and CSS sources, and also to examine any dependence independently on luminosity 
and redshift, one needs different samples to fill in the redshift-luminosity plane.   
With these objectives we observed the Compact Radio sources at Low Redshift (CORALZ) sample, 
which was compiled by Snellen et al. (2004), with the sources
having flux densities larger than 100 mJy at 1400 MHz and angular sizes less than 2 arcsec. 
Snellen et al. (2004) used the 
Faint Images of the Radio Sky at Twenty-Centimeters (FIRST) survey (White et al. 1997), the 
Automated Plate Measuring (APM) machine catalogue (McMahon \& Irwin 1992) of the Palomar 
Observatory Sky Survey (POSS) and their own observations to select radio sources identified
with bright galaxies. Further observations led to a sample
of 18 sources with redshifts in the range 0.005$<$z$<$0.16 which they estimate to be $\sim$95 per
cent complete (Snellen et al. 2004; de Vries et al. 2009).  
Almost all the sources are CSS or GPS objects, suitable for studying young radio sources in the 
nearby Universe. These sources are also significantly weaker than the CSS or GPS objects 
studied so far (e.g. Gupta et al. 2006 and references therein) making it possible to determine gas 
properties in sources of low radio luminosity. The median 5 GHz radio luminosity of these sources 
is $\sim$120 times weaker than those studied by Gupta et al. (2006).

In this paper, we present the results of our observations with the Giant Metrewave Radio Telescope (GMRT) of the 
CORALZ sample. The observations and data analyses are described in Section 2, while the observational results
are described in Section 3. The results are discussed in Section 4, and summarized in Section 5.

\section{Observations and Data Analysis}
These observations were made with the GMRT during 2009 December to 2010 February, using a base-band
bandwidth of 4 MHz ($\sim$900 km s$^{-1}$) in the hardware correlator and an integration time of 16s.  
Each source was observed
for $\sim$4 to 5 hr including calibration overheads. The 4 MHz 
bandwidth consisted of 128 spectral channels, giving a spectral resolution of $\sim$7 km s$^{-1}$. 
Simultaneously data were also recorded using the 
software correlator with a wider bandwidth of 16 MHz over 256 channels and integration times of 16s as well 
as 2s. The data were acquired with a small visibility integration time from the correlator 
to help in the identification of radio frequency interference (RFI). 
The frequencies were tuned such that the expected H{\sc i} absorption line for each source,
corresponding to the optical redshifts listed by Snellen et al. (2004) and the NASA Extragalactic
Database (NED), were within the observing bands. The flux density and bandpass calibrators were either
3C48, 3C147 or 3C286, and  phase calibrators were observed before and after each scan of a source. 
Observational details are listed in Table~\ref{obslog}. 
The data reduction was mainly done using {\tt AIPS++}. The data were reduced using an automated pipeline developed by one of us (SS). 
After applying bandpass corrections on the phase calibrators, gain and phase variations were estimated.
The flux density, bandpass, gain and phase calibration from calibrators were applied to the target sources.

While calibrating the data, bad data were flagged at various stages.
The data for antennas with high errors ($7\sigma$) in antenna-based solutions
were examined and flagged over certain time ranges. Some baselines
were flagged based on closure errors on the bandpass calibrator.
Channel and time-based flagging of data points corrupted by
RFI were done using a median filter
with a $6\sigma$ threshold. Residual errors above $5\sigma$ were
also flagged after a few rounds of target field calibration (using point
source model). Both polarizations (RR and LL) were processed independently 
for consistency checks. The linear fits (of line free channels) at an interval 
of 5 minutes were subtracted from the calibrated data. 
The spectra for the targets were made after averaging the resulting data and correcting for 
the velocities for the Earth's motion using dopset.
Data from both the software and hardware correlator were analysed for 
consistency checks. The spectra were similar and the ones presented 
here are from the hardware correlator because it has twice
the spectral resolution than those obtained from the  software correlator data.
The typical channel r.m.s. for the spectra presented here are $\sim$1.3 mJy. 

\begin {table*}
\caption {Observational details of the GMRT search for H{\sc i} 
          absorption in the CORALZ sample.}
\begin {center}
     \begin {tabular}{ c c c c c c}
     \hline
      Source        & Date       & COB$^1$ &  Total    &Flux density  &  Phase Calibrator    \\
      Name          &            & Freq    & Observation time & and Bandpass &                      \\
                    &            & MHz     &  hr           & Calibrator   &                       \\
     \hline
      J073328+560541&2009 Dec 31 &  1286.5& 5              & 3C147      & J0713+438         \\
      J073934+495438&2010 Jan 05 &  1347.5& 5              & 3C286      & J0713+438         \\ 
      J083139+460800&2010 Jan 05 &  1260.0& 5              & 3C286	& J0713+438         \\
      J083637+440109&2010 Jan 04 &  1346.0& 4              & 3C286 	& J0713+438         \\
      J090615+463618&2010 Feb 03 &  1309.0& 5              & 3C286      & J0713+438         \\
      J102618+454229&2010 Jan 02 &  1232.0& 5              & 3C147      & J1219+484        \\
      J103719+433515&2009 Dec 29 &  1388.5& 5              & 3C286 	& J1219+484         \\
      J115000+552821&2010 Jan 03 &  1247.3& 5              & 3C286      & 3C286         \\
      J120902+411559&2010 Jan 01 &  1297.0& 5              & 3C147      & J1227+365         \\
      J131739+411545&2009 Dec 29 &  1332.5& 5              & 3C286 	& J1227+365         \\
      J140051+521606&2010 Jan 27 &  1270.5& 5              & 3C286      & 3C286         \\
      J140942+360416&2010 Jan 03 &  1237.0& 5              & 3C286      & 3C286         \\
      J143521+505122&2010 Jan 02 &  1292.5& 5              & 3C286      & 3C286        \\
      J150805+342323&2010 Jan 01 &  1359.0& 5              & 3C286 	& 3C286         \\
      J160246+524358&2010 Jan 04 &  1284.5& 4              & 3C286	& J1634+627         \\
      J161148+404020&2010 Feb 09 &  1233.0& 5              & 3C286      & J1613+342         \\
      J170330+454047&2009 Dec 11 &  1340.0& 5              & 3C286      & J1613+342         \\
      J171854+544148&2010 Feb 02 &  1238.5& 5              & 3C48       & J1634+627         \\

    \hline
    \end {tabular}
\end {center}
\label{obslog}
$^1$  COB Freq: Centre of Band Frequency before any correction for velocity using dopset
\end {table*}
  
\section{Observational Results}
\subsection {H{\sc i} observations}
Some of the basic properties of the sample of sources and the observational 
results are summarized in Table~\ref{sourcechar}, which is largely self explanatory. 
The classification of GPS and CSS sources are based on the spectra presented
by Snellen et al. (2004) and any additional information from Labiano et al. (2007).
Sources with a reasonably well-defined spectral peak at frequencies $\gapp$500 MHz
have been classified as GPS, while the others have been termed as CSS objects.
J102618+454618 has been classified as CFS (Compact Flat Spectrum) because it appears to have a 
high-frequency spectral index of $\sim$0.41 (Snellen et al. 2004; de Vries et al. 2009).
From observations of the sources in the CORALZ sample with the GMRT, we report the 
detection of H{\sc i} absorption in 7 sources, of which 3 are GPS sources and 4 are
CSS objects (Table~\ref{sourcechar}). 
The detected absorption lines were fitted with multiple Gaussian components to determine
the peak optical depth $\tau_{p}$ and FWHM ($\Delta v$; km s$^{-1}$) of the spectral components.
H{\sc i} column densities were determined using the relation 
\begin{eqnarray}
{\rm N(HI)} & = & 1.835\times 10^{18}\frac{{\rm T_s}\int\!\tau (v)\,dv}{f_c} \, cm^{-2}  \nonumber \\
  & = &  1.93\times10^{18}\frac{{\rm T_s}\tau_p\Delta v}{f_c}  \, cm^{-2}
\label{eqcol}
\end{eqnarray}
where $T_{s}$ and $f_{c}$ are the spin temperature (in K) and the fraction of the background source
covered by an absorbing cloud. We have assumed $T_{s}$=100 K and $f_{c}$=1.0. The column densities
range from $\sim$1.78$\times$10$^{20}$ to $\sim$10$^{22}$ cm$^{-2}$, with a median value of 
$\sim$7$\times$10$^{20}$ cm$^{-2}$. These values are similar to those of the more 
luminous GPS and CSS sources where the median value is $\sim$5$\times$10$^{20}$cm$^{-2}$ (Gupta et al. 2006). 

\begin{figure*}
\vbox{
\hbox{
\psfig{figure=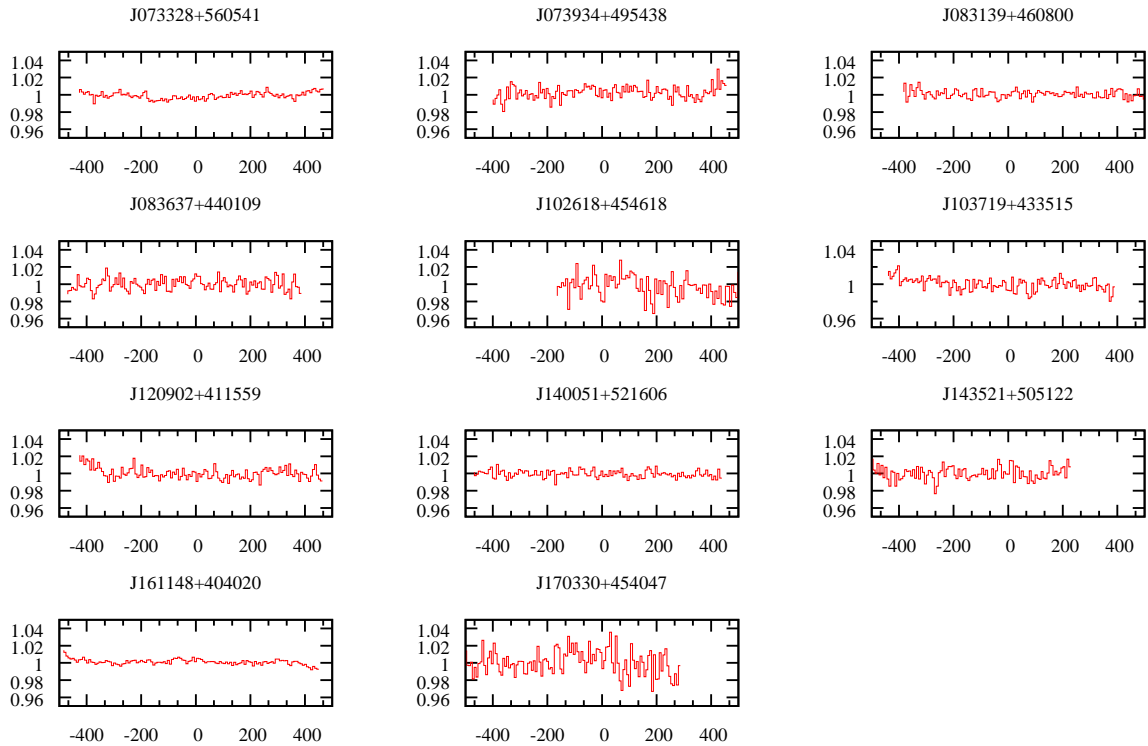,width= 18 cm,angle=-90}
}
}
\vspace{-2cm}
\caption[]{GMRT spectra for the non-detections. The y-axis shows the normalised intensity while the 
           x-axis shows the velocity in km s$^{-1}$.} 
\label{ND}
\end{figure*}

\begin{figure*}
\vspace{-0.35cm}
\hspace{-0.3cm}
\vbox{
\hbox{
\psfig{figure=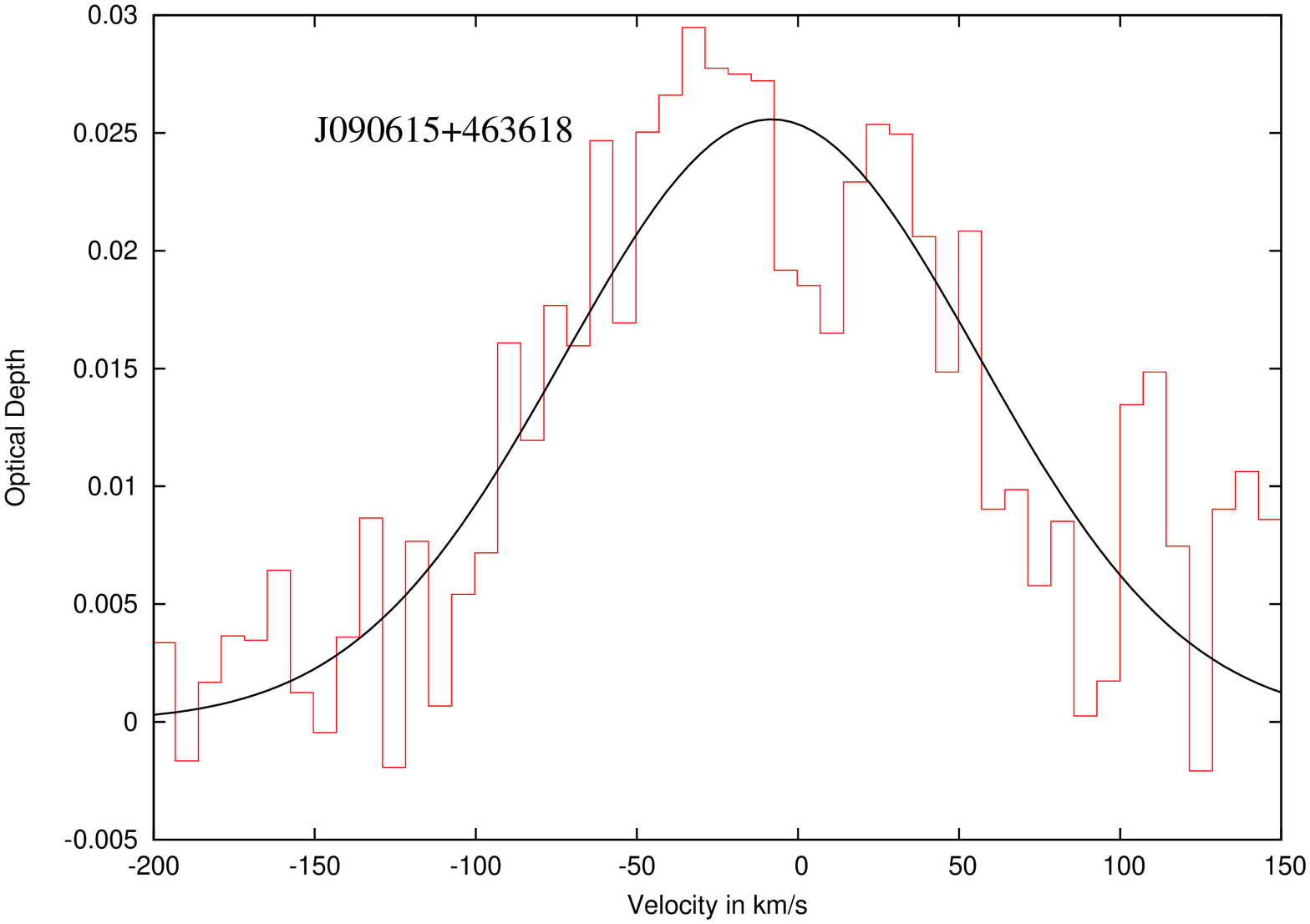,width=8.4cm,angle=-90}
\psfig{figure=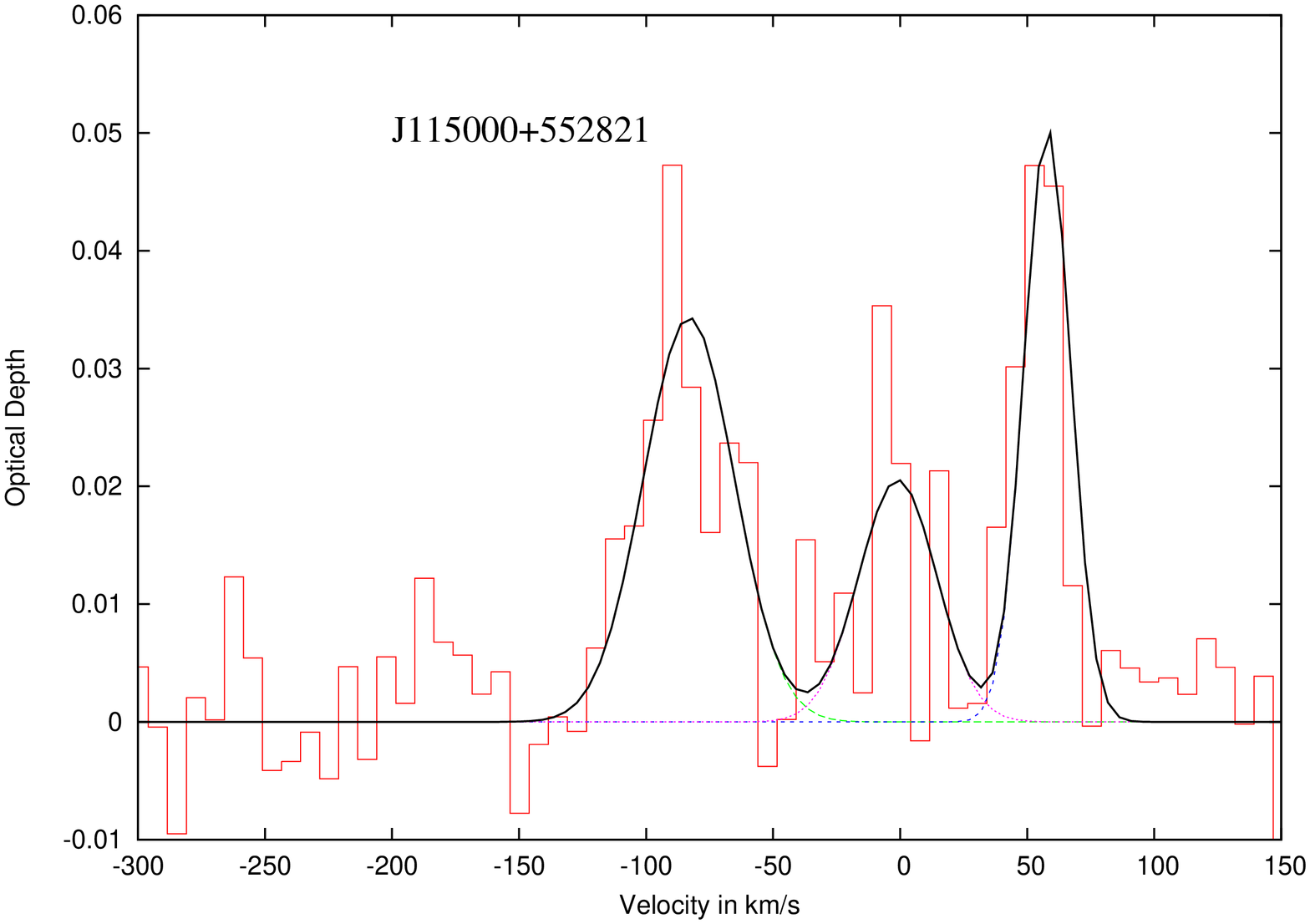,width=8.4cm,angle=-90}
}
\hbox{
\psfig{figure=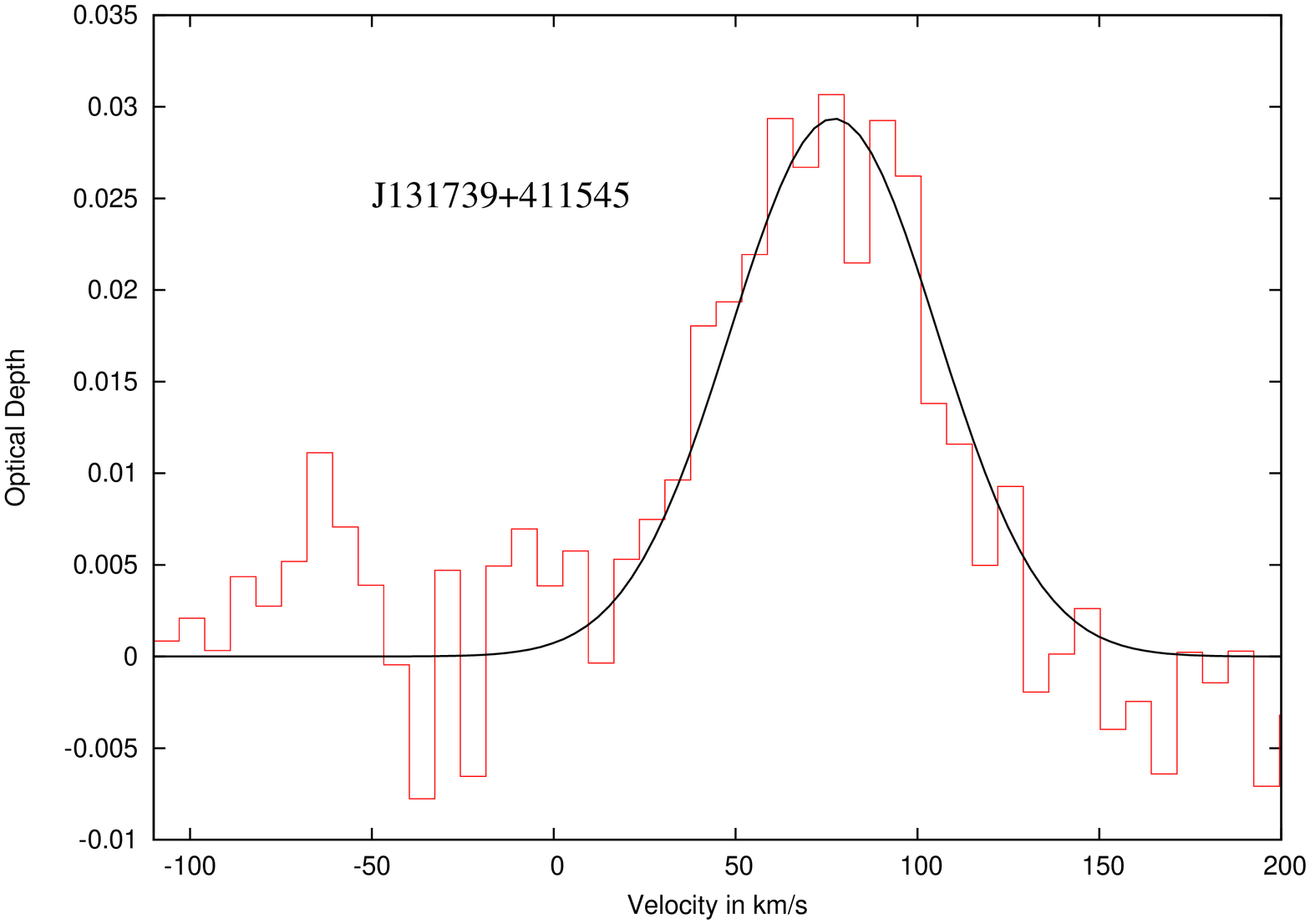,width=8.4cm,angle=-90}
\psfig{figure=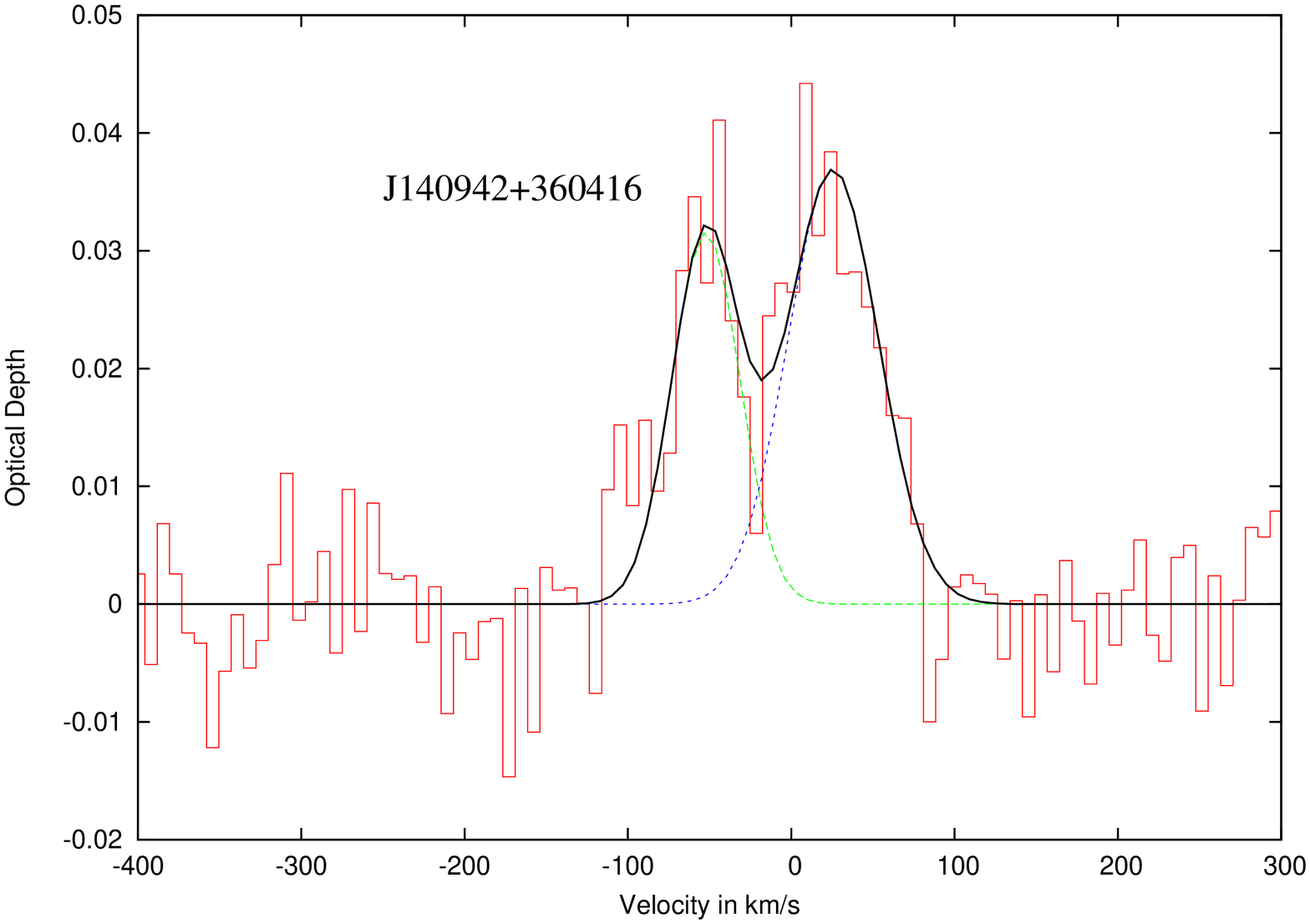,width=8.4cm,angle=-90}
}

\hbox{
\psfig{figure=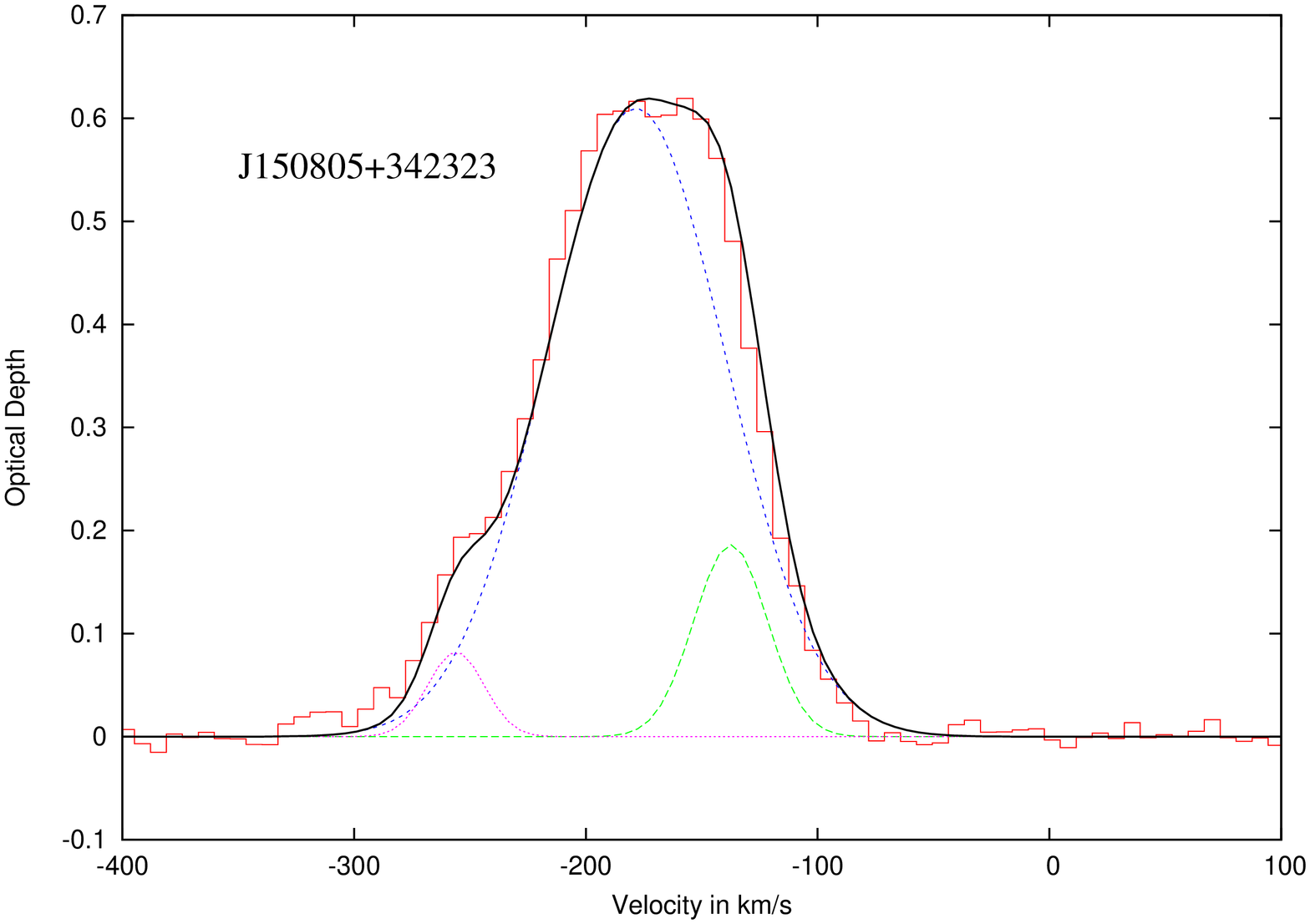,width=8.4cm,angle=-90}
\psfig{figure=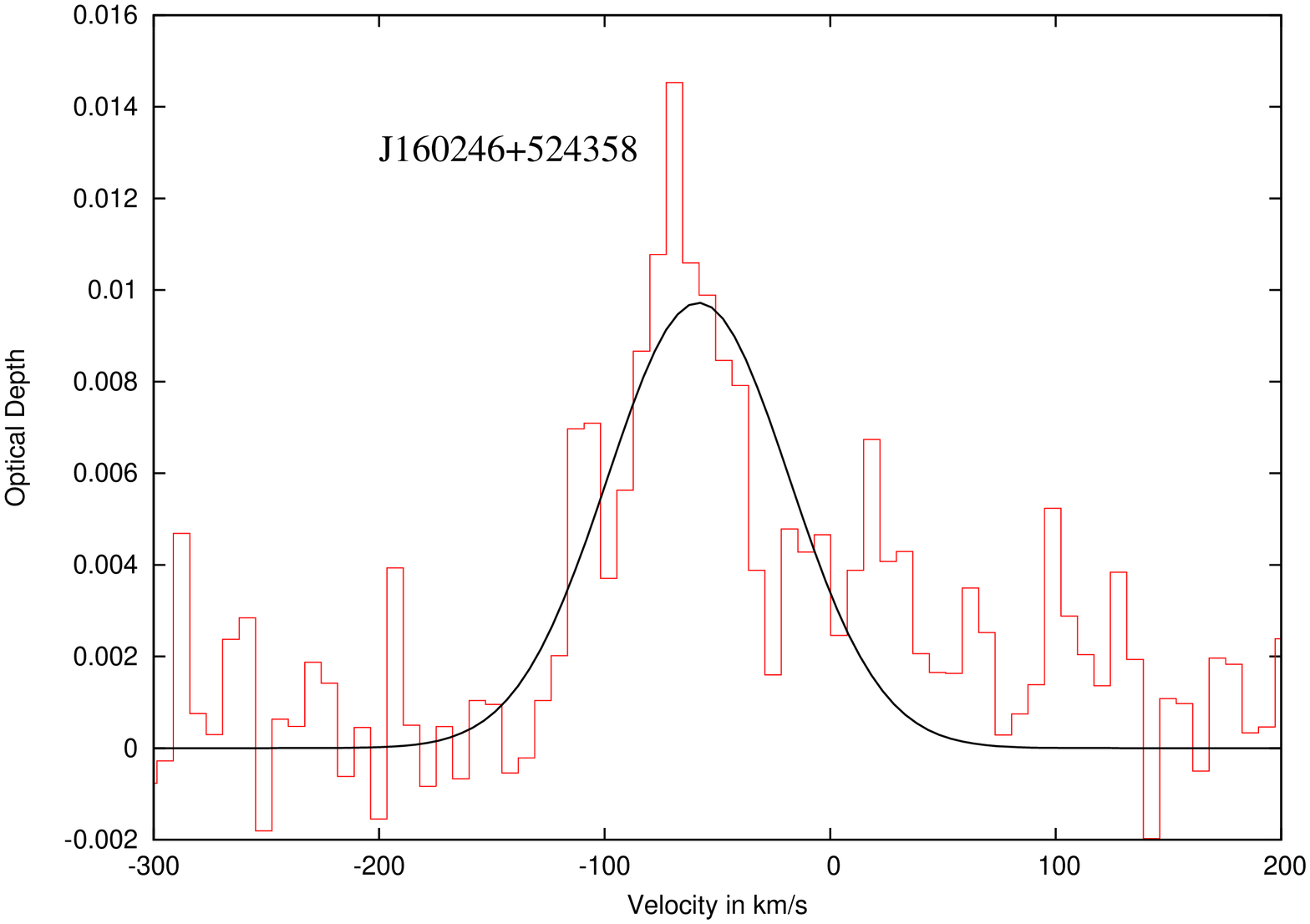,width=8.4cm,angle=-90}
}
\hbox{
\psfig{figure=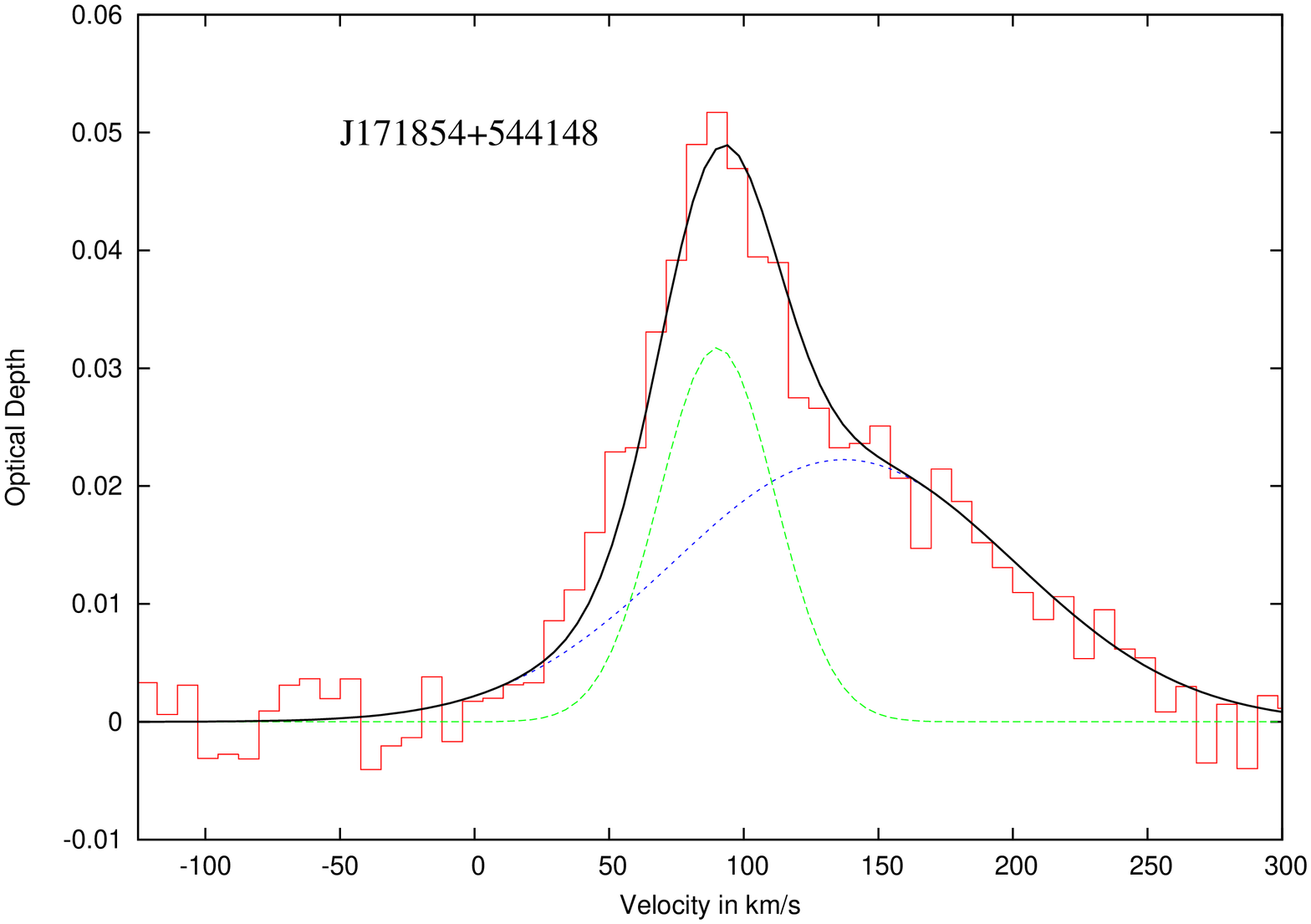,width=8.4cm,angle=-90}
}
}
\vspace{-0.35cm}
\caption[]{The optical depth of the H{\sc i} absorbers vs 
           velocity in km s$^{-1}$ relative to the systemic velocity.}
\label{GF1}
\end{figure*}

For the non-detections, we have determined the upper limits on H{\sc i} column densities using 
3$\times$r.m.s. of the optical depths as $\tau_{p}$ 
and $\Delta v$=100 km s$^{-1}$. These upper limits range from $\sim$0.9 to $\sim$4.2$\times$10$^{20}$ cm$^{-2}$.
The spectra for the non-detections are shown in Fig.~\ref{ND}, while the spectra for the 
detections are presented in Fig.~\ref{GF1} along with the Gaussian fits to the
optical depths.  All the spectra have been plotted relative to the systemic velocity inferred
from the redshifts listed in Table 2, which are from NED, 
but taken mainly from the Sloan Digital Sky Survey (SDSS; Adelman-McCarthy et al. 2008, and references therein), 
except for J083139+460800 and J103719+433515
which have been plotted relative to the redshifts listed by Snellen et al. (2004). For all the non-detections,
the absence of H{\sc i} absorption has also been confirmed from the wider bandwidth software correlator data, 
except for J103719+433515, where there was an error in the settings. 
We describe below each source briefly including  J083139+460800 and J103719+433515, which is then
followed by the discussion Section.

\begin {table*}
\caption {Characteristics of sources in the CORALZ sample. The redshifts, $z$, are from NED with the original
          references being listed in Column 4, while the flux densities
          are from the FIRST survey. The luminosities from Snellen et al. (2004) have been re-estimated in the 
          cosmology used here. The largest angular sizes ($\theta$) are from de Vries et al. (2009) 
          except for J073328+560541 (Bondi et al. 2001) and  J170330+454047 (Gu \& Chen 2010). 
          LS denotes the corresponding linear sizes.}   
\begin {center}
     \begin {tabular}{c c c c c c r r c r}
     \hline
     Source         & Opt  & $z$   & Refs. &S$_{1.4GHz}$&   $L_{5GHz}$         & $\theta$& LS    & Spectral&  N(H{\sc i})\\
     Name           & Id.  &       &       &    mJy     &10$^{25}$(W $Hz^{-1}$)& (mas)   &(pc)   & Class   &  $10^{20}$ cm$^{-2}$\\
         (1)        &  (2) & (3)   &  (4)  &   (5)      &    (6)               &   (7)   & (8)   &  (9)    &   (10)        \\
     \hline
     J073328+560541 &   G  & 0.1040 & 1    &   348      &  0.467               &  80   & 151     & GPS     &         $<$1.076  \\
     J073934+495438 &   G  & 0.0540 & 1,2  &   107      &  0.042               &$<$2   &$<$2.1   & GPS     &         $<$2.097  \\
     J083139+460800 &   G  & 0.1311 & 3    &   131      &  0.408               &   9   &  20.7   & GPS     &         $<$1.256   \\
     J083637+440109 &   G  & 0.0554 & 3    &   139      &  0.045               & 1600  & 1699    & CSS     &         $<$1.945   \\
     J090615+463618 &   G  & 0.0848 & 1,3  &   314      &  0.302               &  31   & 48.9    & GPS     &            7.482  \\
     J102618+454618 &   G  & 0.1517 & 3    &   105      &  0.347               &  17   & 44.7    & CFS     &         $<$3.617  \\
     J103719+433515 &   G  & 0.0247 & 3,4  &   129      &  0.009               &  19   & 8.7     & CSS     &         $<$1.932 \\
     J115000+552821 &   G  & 0.1385 & 3    &   143      &  0.363               &  41   & 93.9    & CSS     &            6.31  \\
     J120902+411559 &   G  & 0.0950 & 2    &   147      &  0.178               &  20   & 34.8    & CSS     &         $<$1.854 \\
     J131739+411545 &   G  & 0.0662 & 1,3  &   249      &  0.229               &   4   & 5       & GPS     &          3.785   \\
     J140051+521606 &   G  & 0.1180 & 3    &   174      &  0.224               &$<$150 &$<$316   & CSS     &         $<$1.139  \\
     J140942+360416 &   G  & 0.1484 & 3    &   143      &  0.276               &  27   & 69.2    & CSS     &          7.667    \\
     J143521+505122 &   G  & 0.0997 & 3    &   141      &  0.155               &$<$150 &$<$271   & CSS     &         $<$2.339 \\
     J150805+342323 &   G  & 0.0456 & 5    &   130      &  0.022               & 170   & 148     & CSS     &         125.183 \\
     J160246+524358 &   G  & 0.1057 & 3    &   576      &  0.549               & 180   &345      & CSS     &         1.781\\
     J161148+404020 &   G  & 0.1520 & 2    &   553      &  1.048               & 1300  &3400     & CSS     &         $<$0.929  \\
     J170330+454047 &   G  & 0.0604 & 6    &   119      &  0.034               &$<$7.3 &$<$8.4   & CSS     &         $<$4.165   \\
     J171854+544148 &   G  & 0.1470 & 1,7  &   329      &  0.617               & 68    & 172.9   & GPS     &          9.541    \\
     \hline
     \end {tabular}

References for redshifts: 1 Labiano et al. (2007); 2 Snellen et al. (2004); 3 SDSS (Adelman-McCarthy et al. 2008 and references therein);
            4 Falco et al. (1999); 5 Mazzarella et al. (1993); 6 de Grijp et al. (1992); 7 Kim \& Sanders (1998)
\end {center}
\label{sourcechar}
 \end {table*}

\subsection{Notes on individual sources}
\noindent
{\bf J073328+560541:} Very Long Baseline Interferometric (VLBI) images at $\sim$1.6 and 5 GHz 
show the source to have an extent of $\sim$80 mas (Bondi et al. 2001; de Vries et al. 2009), while Very Large Array 
A-array observations at 8.4 GHz show an unresolved source with an angular extent $<$0.2 arcsec 
(Patnaik et al. 1992). The location of the core is unclear and the source is unpolarized (Bondi et al. 2004).
Goncalves \& Serote Roos (2004) suggest the presence of Low Ionization Nuclear
Emission-line Region (LINER), while Dennett-Thorpe \& March\~a (2000) suggest that
the source might be variable at 1.4 GHz. The radio spectrum peaks at $\sim$460 MHz
(Snellen et al. 2004). We do not find evidence of H{\sc i} absorption. 

\noindent
{\bf J073934+495438:}  VLBI images of the source at $\sim$5 GHz show it to be unresolved
(de Vries et al. 2009). The radio spectrum peaks at $\sim$950 MHz (Snellen et al. 2004).
We do not find evidence of H{\sc i} absorption.

\noindent
{\bf J083139+460800:} VLBI observations show the source to be double lobed with an overall
separation of 4.4 mas in the image of epoch 2000, while a weak component visible towards
the north in the image of epoch 2004 gives an overall size of $\sim$9 mas (de Vries et al. 2009). 
The redshift of the host galaxy from SDSS is 0.1311, while Snellen et al. (2004)
list it as 0.127.  It is a GPS source with a turnover at $\sim$2200 MHz (Snellen et al. 2004). We 
do not find evidence of H{\sc i} 
absorption. In the spectrum shown in Fig.~\ref{ND}, zero velocity corresponds to 
optical redshift of 0.127 listed by Snellen et al. (2004) and de Vries et al. (2009). While the 
redshift of 0.1311 is not within the range covered by the hardware correlator, it is within the range
covered by the software correlator and no H{\sc i} absorption is seen. 

\noindent
{\bf J083637+440109:} The MERLIN image at 1.6 GHz (de Vries et al. 2009) shows most of 
the emission to form a C-shaped structure with an angular size of $\sim$1600 mas. 
We do not find evidence of H{\sc i} absorption. 

\noindent
{\bf J090615+463618:} The radio structure revealed by the VLBI maps 
(de Vries et al. 2009; Helmboldt et al. 2007) 
has been suggested to be due to emission on opposite sides of the core by 
Bondi et al. (2004). 
The source exhibits no significant polarization at both radio ($<$0.99$\%$) 
and optical ($<$1.2$\%$) wavelengths (March\~a et al. 1996; Dennett-Thorpe \& March\~a 2000). 
Caccianiga et al. (2002) classify it as a Narrow Emission Line Galaxy, while Goncalves \& Serote Roos (2004) 
suggest it to be a LINER. We detect H{\sc i} absorption in this source with one main 
component blueshifted by $\sim$8 
km s$^{-1}$ (Fig.~\ref{GF1}) for which the Gaussian fit parameters are given in Table.~\ref{D1}.  

\noindent
{\bf J102618+454618:} The VLBI map at 5 GHz (de Vries et al. 2009), has multiple components
with an overall separation of 17 mas, with the central feature being the most prominent one.
From the VLBI maps at 1665 and 4993 MHz (de Vries et al. 2009), the central feature  
has a flat spectra. We do not find evidence of H{\sc i} absorption.


\begin{table*}
\caption {Multiple Gaussian fit parameters to the H{\sc i} absorption spectra for the detections. }
\begin {center}
     \begin {tabular}{c c c c c c }
     \hline
Object & Component & Vel  &   FWHM & $\tau_{p}$ & N(H{\sc i}) \\
       &  Number & km s$^{-1}$ &  km s$^{-1}$ &    &  $10^{20}$ cm$^{-2}$ \\ 
      \hline

 J090615+463618 & 1   & $-$8.2(4.2)    & 151.5(10.1)     & 0.0256(0.0015) &  7.48(0.93) \\
                &     &                  &                &                &             \\ 
J115000+552821  &  1  & $-$83.1(2.6) & 42.1(6.1)   & 0.0343(0.0043) &  2.79(0.76) \\
                &  2  & $-$0.9(4.0)  & 36.0(9.6)   & 0.0206(0.0047) &  1.43(0.71) \\
                &  3  &   57.9(1.3)  & 21.5(3.0)   & 0.0504(0.0060) &  2.09(0.54) \\
                &     &                  &                &                &             \\ 
J131739+411545  &  1  & 77.0(2.4) & 66.8(5.8)  & 0.0294(0.0022) & 3.79(0.61) \\
                &     &                  &                &                &             \\ 
J140942+360416  &1    & $-$52.2(3.4) & 49.2(7.9)  & 0.0316(0.0033) &  3.00(0.80)  \\
                &2    &  25.7(3.3)   & 65.6(8.4)  & 0.0369(0.0029) &  4.67(0.97)  \\
                &     &                  &                &                &             \\ 
J150805+342323  & 1   & $-$256.1(1.4) &  28.8(3.6) & 0.0822(0.0083) &  4.56(1.04)  \\
                &2    & $-$178.4(1.1) &  91.1(1.7) & 0.6100(0.0084) &  107.20(3.46)  \\
                &3    & $-$137.5(0.9) &  37.4(3.1) & 0.1862(0.0178) &  13.42(2.39)  \\
                &     &                  &                &                &             \\ 
J160246+524358  &1    &  $-$58.4(4.2) & 94.9(9.8) & 0.0097(0.0009)  & 1.78(0.34) \\
                &     &                  &                &                &             \\ 
J171854+544148  &1    &  90.1(1.4) & 50.6(4.2)   & 0.0317(0.0024) &  3.10(0.49) \\
                &2    & 137.2(5.7) & 149.9(10.9) & 0.0223(0.0019) &  6.44(1.01) \\
\hline
     \end {tabular}
\end {center}
\label{D1}
\end {table*}



\noindent
{\bf J103719+433515:} This source too has an elongated structure with multiple components and 
an overall size of 19 mas. The brightest component in the central region of the source appears to
have a flat spectrum (de Vries et al. 2009). 
There is no evidence of H{\sc i} absorption. The zero velocity in the spectrum
corresponds to the optical redshift of 0.023 as given by Snellen et al. (2004), while the one given by 
NED is 0.0247 (Falco et al. 1999), which was not covered by the hardware correlator. There was also an error in the 
settings of the software correlator. The source needs to be re-observed to ensure that there is also 
no absorption at a redshift of 0.0247. For the present, we have classified it as a non-detection.

\noindent
{\bf J115000+552821:} The VLBI map at 1.6 GHz shows a dominant compact component
with weak emission towards the east separated from it by $\sim$41 mas (de Vries et al. 2009).
We detect H{\sc i} absorption, with three components (Fig.~\ref{GF1}), 
one of which is redshifted and one blueshifted relative to the systemic velocity. The 
third one, which is weaker and would be useful to confirm, is consistent with the systemic velocity. 
The Gaussian fitting parameters are given in Table.~\ref{D1}.

\noindent
{\bf J120902+411559:} The source exhibits multiple components with a prominent central
feature and is elongated along the east-west direction with angular size $\sim$20 mas
(de Vries et al. 2009).  The radio spectrum of this source peaks at 370 MHz (Snellen et al. 2004). 
There is no evidence of H{\sc i} absorption. 

\noindent
{\bf J131739+411545:}  VLBI maps show a complex structure with size of $\sim$4 mas
(Helmboldt et al. 2007; de Vries et al. 2009). The radio spectrum of this source peaks at
2.3 GHz (Snellen et al. 2004). We detect H{\sc i} absorption, where the 
line profile is redshifted with respect to the optical redshift by 77 km s$^{-1}$
indicating infall of gas towards the central source (Fig.~\ref{GF1}). The Gaussian fit
parameters are listed in Table~\ref{D1}.

\noindent
{\bf J140051+521606:} The MERLIN map at 1.4 GHz shows a resolved component with a deconvolved angular
size of 182.3$\times$95.9 mas along a PA of 102$^\circ$ (Table B2 of de Vries et al. 2009).
There is no evidence of H{\sc i} absorption.

\noindent 
{\bf J140942+360416:} The VLBI images at $\sim$1.7 GHz suggest that it could be a very asymmetric double.
The higher resolution image in  de Vries et al. (2009) shows the component towards the north to have a
weaker flux density by a factor of $\sim$10. The radio spectrum of this source peaks at 330 MHz (Snellen et al. 2004). 
We detect H{\sc i} absorption towards this source. The H{\sc i} line profile shows two components, one blueshifted while the 
second one is redshifted relative to the systemic velocity, consistent with rotation of the absorbing gas
(Fig.~\ref{GF1}; Table~\ref{D1}).  

\noindent
{\bf J143521+505122:} The MERLIN image at 1.4 GHz shows extended emission 
which has been deconvolved into two well-resolved components (de Vries et al. 2009). 
There is no detection of H{\sc i} absorption.

\noindent
{\bf J150805+342323:} The VLBI map at 1.6 GHz shows two components.
(de Vries et al. 2009). CO emission has been detected from this source (Mazzarella et al. 1993;
Evans et al. 2005). There is evidence of strong H{\sc i} absorption 
(Fig.~\ref{GF1}; Table~\ref{D1}). The Gaussian fit to the optical depth and 
velocity shows blueshifted gas with three main components which could be due 
to outflowing gas.

\noindent
{\bf J160246+524358:} The VLBI map at 5 GHz (de Vries et al. 2009) shows a 
compact flat-spectrum core with jet-like emission on opposite sides. 
The optical source has been classified as a Seyfert 1 AGN by V\'eron \& V\'eron (2006). There is evidence of blueshifted 
H{\sc i} absorbing gas (Fig~\ref{GF1}; Table~\ref{D1}).     

\noindent
{\bf J161148+404020:} The complex structure of this source is seen in a
MERLIN map at 1.6 GHz with an overall angular size of 1300 milliarcsec (de Vries et al. 2009). 
There is no detection of H{\sc i} absorption. 


\begin{figure}
\vbox{
\psfig{figure=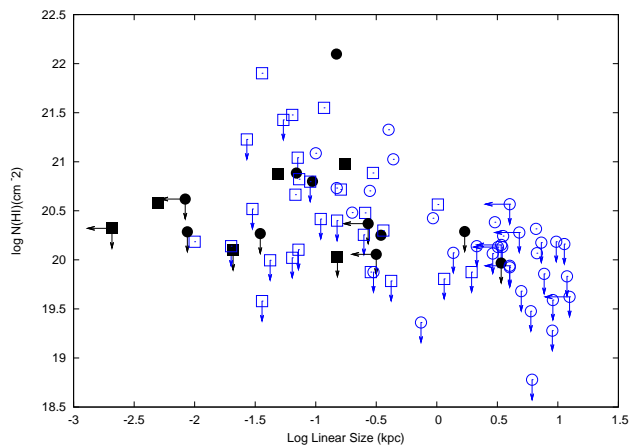,width=8.5cm,angle=-90}
     }
\caption[]{H{\sc i} column density versus projected linear size for the CORALZ sample
as well as the higher luminosity CSS and GPS objects. The black filled symbols denote the
CORALZ sources, while blue open symbols denote the sources from the Gupta et al. (2006) compilation.
CSS and GPS sources are represented by circles and squares  respectively.
}
\label{ColD_size}
\end{figure}


\begin{figure}
\vbox{
\psfig{figure=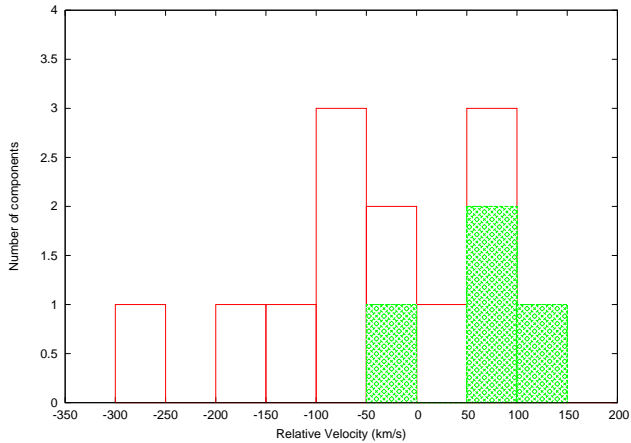,width=8.5cm,angle=-90}
}
\caption[]{The distribution of the velocity of the different H{\sc i} absorbing components
relative to the systemic velocities of the host galaxies. The GPS sources are shown shaded.} 
\label{vel_diff}
\end{figure}

\noindent 
{\bf J170330+454047:}  This is a compact unresolved radio source as observed in the Japanese VLBI Network map 
at 8.4 GHz (Doi et al. 2007), while VLBA map at 5 GHz shows two components resolved into core-jet (Gu \& Chen 2010). 
This source was classified originally as Seyfert 2 (de Grijp et al. 1992) but later on reclassified as a 
Narrow Line Seyfert 1 (Moran et al. 1996; Wisotzki \& Bade 1997). 
There is no evidence of H{\sc i} absorption.

\noindent
{\bf J171854+544148:} This source shows two prominent components with a separation of $\sim$ 68 mas at
$\sim$1.6 GHz, reminiscent of a CSO, but shows two weaker components in a higher-resolution image
at $\sim$5 GHz (de Vries et al. 2009). This source has been classified as a Seyfert type 2 
(Veilleux et al. 1999). It was earlier classified by Leech et al. (1994) as an interacting system,
but the second object appears to be a foreground star (Bruston, Ward \& Davies 2001; Davies, 
Burston \&  Ward 2002). The object is also associated with an ultraluminous infrared galaxy (ULIRG) 
F17179+5444 (Leech et al. 2004). The radio spectrum of the source peaks at $\sim$480 MHz. 
We detect H{\sc i} absorption towards this source (Fig.~\ref{GF1}; Table~\ref{D1}), with two
main components, both of which are redshifted relative to the systemic velocity.
 
\section{Discussion}
The radio luminosity of the sources in the core CORALZ sample which have a flux density at
1400 MHz $>$100 mJy, redshift in the range 0.005 to 0.16 and angular size $<$2 arcsec (Snellen et al. 2004;
de Vries et al. 2009) are almost always less than $\sim$10$^{25}$ W Hz$^{-1}$ at 5 GHz. Since the radio luminosities of 
95 per cent of the  sources in the sample compiled by Gupta et al. (2006) are greater than $\sim$10$^{25}$ W Hz$^{-1}$ at 5 GHz,
we have compared our results with those of Gupta et al. (2006) to examine any broad dependence on radio luminosity. 
 In this section, we leave out J102618+454618 which has a flat high-frequency spectrum (Snellen et al. 2004).
In the remaining sample of  17 compact low redshift sources, we were able to detect 7 new sources in 
H{\sc i} absorption.  Of these seven, three are GPS and four are CSS sources, the fraction of detection for
the GPS sources being $\sim$50 per cent (3/6) compared with $\sim$36 per cent (4/11) for CSS objects. Although
the numbers are small and statistical uncertainties are large, these are consistent with the detection rates
for the more luminous sources, and the tendency for GPS objects to have a higher detection rate compared with 
CSS objects (e.g. Vermeulen et al. 2003; Philstr\"om et al. 2003; Gupta et al. 2006). As mentioned earlier in Section 3.1,
the column densities are similar to those for the more luminous CSS and GPS objects.


\begin{figure*}
\vspace{0.6cm}
\vbox{
\centerline{
\psfig{figure=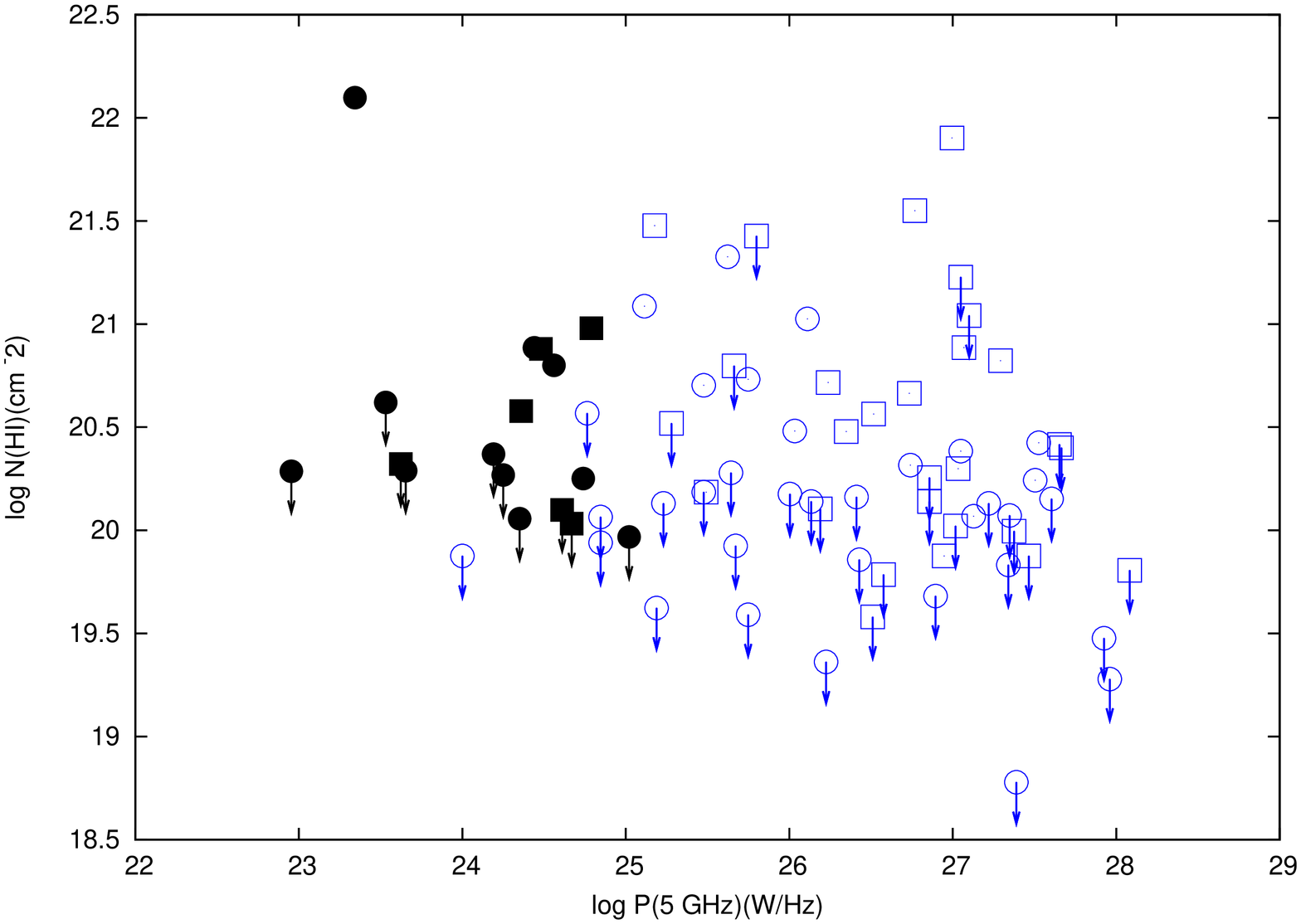,width=8.5cm,angle=-90}
\psfig{figure=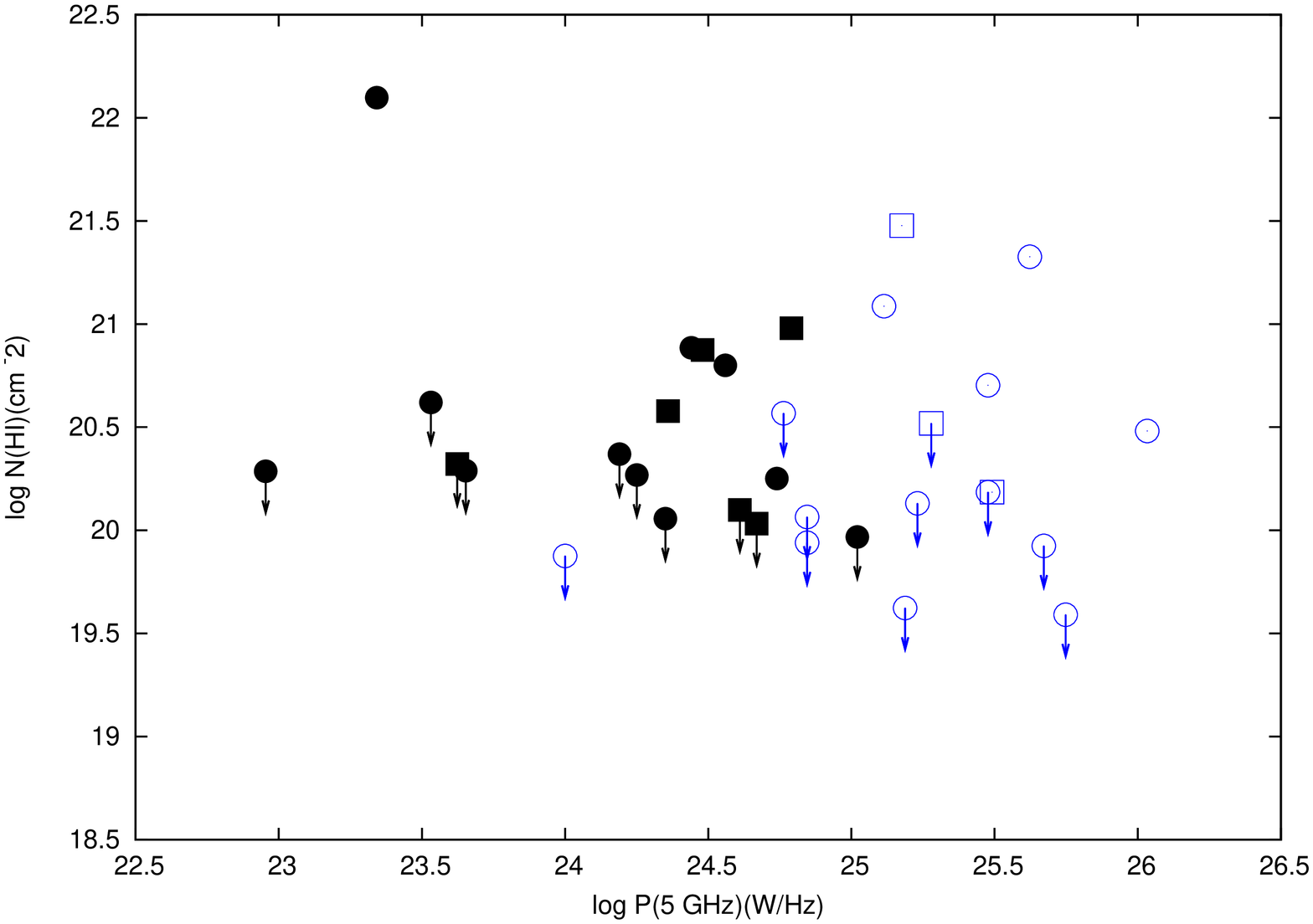,width=8.5cm,angle=-90}
\hspace{0.3cm}
}
}
\caption[]{Left panel: H{\sc i } column density versus source luminosity at 5 GHz for 
           the CORALZ sample (black filled ones) and from the Gupta
           et al. (2006) compilation (blue open ones). The CSS and GPS objects are shown by  circles and squares  respectively.
           Right panel: Same as left but for
           sources with a redshift $<0.2$. 
          }
\label{Lum_ColD1}
\end{figure*}


\begin{figure}
\vbox{
\hbox{
\psfig{figure=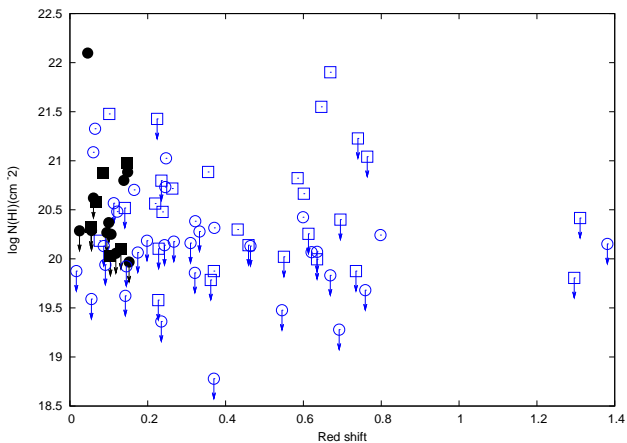,width=8.5cm,angle=-90}
}
}
\caption[]{H{\sc i} Column density versus redshift for the CORALZ sources (black filled symbols) and the 
          CSS and GPS objects from the compilation by Gupta et al. (2006) shown with open blue symbols.
          The CSS and GPS objects are shown by  circles and squares  respectively. 
           }
\label{RS_ColD} \end{figure}


\subsection{H{\sc i} column density vs linear size}
Earlier studies have shown an inverse correlation between the H{\sc i} column density and the projected linear
size. This was first noticed by Pihlstr\"om et al. (2003) and was later shown to be also true for a significantly
larger sample of sources by  Gupta et al. (2006).  In Fig.~\ref{ColD_size} we have plotted the results of our
observations for these 17 sources in the H{\sc i} column density vs linear size diagram, along with the list
of higher-luminosity sources compiled by Gupta et al. (2006). These low-luminosity radio sources from the CORALZ
sample are consistent with a similar relationship, suggesting that the most compact sources are seen through
regions of higher column density. 

\subsection{Relative velocity of H{\sc i} absorption features}
The distribution of the relative velocity for the H{\sc i} absorbing components listed in Table~\ref{D1} 
is shown in Fig.~\ref{vel_diff}. These range from $\sim$$-$250 to 140 km s$^{-1}$, with a median value of 
$\sim$$-$50 km s$^{-1}$. Although the number of absorption features is small, and there are uncertainties in
the systemic velocities (e.g. Morganti et al. 2001) there appears to be a marginal trend for a higher number of
blueshifted features.  

Although blueshifted H{\sc i} absorption features could also arise from either halo or circumnuclear gas
affected by winds and/or radiation pressure from the nuclear region, jet-cloud interactions are likely to 
significantly affect the gas properties in compact steep-spectrum radio sources. The tendency for CSS and GPS
objects to be more asymmetric in location, brightness and polarization of the lobes compared with
the larger objects suggest strong interaction of the jets with the circumnuclear region (e.g. Saikia et al. 
1995, 2001; Arshakian \& Longair 2000; Saikia \& Gupta 2003). Rotation measure (RM) studies also indicate high values
in many compact sources (e.g Mantovani et al. 1994, 2010; O'Dea et al. 1998), often with large asymmetries in the RM
values for the oppositely-directed lobes due to interaction with asymmetrically located clouds of gas (e.g.
Junor et al. 1999). Evidence of interaction is also seen in optical emission and absorption 
line studies (e.g. Gelderman \& Whittle 1994; Chatzichristou et al. 1999; Labiano et al. 2005; Gupta et al. 2005).

Excess of blueshifted absorption features has been reported earlier. For example, for a sample of quasars 
Baker et al. (2002) found a small excess of blueshifted C{\sc iv} absorption features in CSS objects compared with
the larger sources which they attribute to absorbing material being away from the jet axis, whereas 
van Ojik et al. (1997) found an excess of blueshifted Ly$\alpha$ absorption features in their sample of 
small-sized high-redshift radio galaxies which they attribute to absorbing clouds uniformly covering the 
whole source signifying a halo origin. The sample of CSS and GPS objects compiled by Gupta et al.
(2006) also shows a tendency for the H{\sc i} absorption features to be blueshifted with velocities extending to
over 1000 km s$^{-1}$. Although the number of low-luminosity sources needs to be increased, the present observations
suggest that the blueshifted velocities in these low-luminosity sources could be smaller possibly due to the weaker
radio jets in these objects.

\subsection{H{\sc i} column density vs luminosity and redshift}
It is generally believed that AGN activity is triggered by the accretion of matter onto a supermassive blackhole
at the centre of the galaxy. Mergers and interactions of galaxies could facilitate the supply and inflow of gas
into the central regions of these active galaxies leading to both circumnuclear starbursts and fuelling the 
supermassive blackholes (e.g. Sanders et al. 1988; Hopkins et al. 2005). It may therefore be relevant to 
investigate whether the H{\sc i} column density may depend on either radio luminosity or redshift. Gupta et al.
(2006) did not find any evidence of such relationships in their sample of luminous CSS and GPS objects. However,
since most of the sources in their sample were selected from strong-source surveys, luminosity and redshift are
strongly correlated. Constraining the sources from the Gupta et al. (2006) sample to those with redshifts $<$0.2,
we have examined any dependence of column density on luminosity over a similar redshift range but find no evidence
of any significant relationship (Fig.~\ref{Lum_ColD1}). The CORALZ sources are of lower luminosity than the 
Gupta et al. (2006) sample making it difficult to examine variation with redshift for similar luminosity objects.
We have plotted the H{\sc i} column density vs redshift for the CORALZ sources along with all the sources from
the Gupta et al. (2006) sample (Fig.~\ref{RS_ColD}) and find no evidence of any correlation.

\section{Summary}
The results of H{\sc i} absorption measurements towards a sample of
nearby CSS and GPS radio sources, the CORALZ sample, using the GMRT are 
summarized briefly here. These sources are of lower luminosity than earlier
studies of CSS and GPS objects.

(i) We observed a sample of 18 sources and find 7 new detections.
    Within the uncertainties caused by the small sample size, the detection 
    rates are similar to the more luminous objects, with the GPS objects again
    exhibiting a higher detection rate (3/6) than for CSS objects (4/11). 

(ii) The relative velocity of the blueshifted absorption features, which
     may be due to jet-cloud interactions, extend to only $\sim$$-$250 km s$^{-1}$
    for these CORALZ sources compared with values of over $\sim$1000 km s$^{-1}$ for 
    the more luminous CSS and GPS objects. This could be due to the weaker jets in these 
    low-luminosity objects, although this needs confirmation from a larger sample of objects.  
   
(iii) There appears to be no evidence of any dependence of H{\sc i} column density
      on either luminosity or redshift. Examining sources over a similar redshift
      range ($<$0.2) also shows no evidence of any correlation with luminosity. 

(iv) The weaker CSS and GPS objects are also consistent with the 
     known inverse relation of H{\sc i} column density with projected linear size.

\section*{Acknowledgments}
We thank the reviewer for his/her detailed and valuable comments which have helped improved the manuscript 
significantly. We thank the staff of GMRT for their assistance during our observations.
The GMRT is a national facility operated by the National Centre for Radio Astrophysics of the Tata
Institute of Fundamental Research.  We thank  numerous contributors to the GNU/Linux group.
This research has made use of the NASA/IPAC Extragalactic Database (NED) which is operated by the Jet
Propulsion Laboratory, California Institute of Technology, under contract with the National
Aeronautics and Space Administration.  We have made use of the cosmological calculator by Edward L. Wright
(Wright 2006). This research has made use of NASA's Astrophysics Data System.
%

{\small
{}
}

\begin{thebibliography}{}
\bibitem[]{} Arshakian T.G., Longair M.S., 2000, MNRAS, 311, 846
\bibitem[]{} Adelman-McCarthy et al. 2008, ApJS, 175, 297
\bibitem[]{} Baker J.C., Hunstead R.W., Athreya R.M., Barthel P.D., de Silva E., Lehnert M.D., Saunders R.D.E., 2002, ApJ, 568, 592
\bibitem[]{} Bondi M., March\~a M.J.M., Dallacasa D., Stanghellini C., 2001, MNRAS, 325, 1109
\bibitem[]{} Bondi M., March\~a M.J.M., Polatidis A., Dallacasa D., Stanghellini C., Ant\'on S., 2004, MNRAS, 352, 112
\bibitem[]{} Bruston A.J., Ward M.J., Davies R.I., 2001, MNRAS, 326, 403
\bibitem[]{} Caccianiga A., March\~a M.J., Ant\'on S., Mack K.-H., Neeser M.J., 2002, MNRAS, 329, 877
\bibitem[]{} Chatzichristou E.T., Vanderriest C., Jaffe W., 1999, A\&A, 343, 407
\bibitem[]{} Davies R.I., Burston A.J., Ward M.J., 2002, MNRAS, 329, 367
\bibitem[]{} de Grijp M.H.K., Keel W.C., Miley G.K., Goudfrooij P., Lub J., 1992, A\&AS, 96, 389
\bibitem[]{} Dennett-Thorpe J., March\~a M.J., 2000, A\&A, 361, 480
\bibitem[]{} Doi A., et al., 2007, PASJ, 59, 703
\bibitem[]{} de Vries N., Snellen I.A.G., Schilizzi R.T., Mack K.H., Kaiser C.R., 2009, A\&A, 498, 641
\bibitem[]{} Evans A.S., Mazzarella J.M., Surace J.A., Frayer D.T., Iwasawa K., Sanders D.B., 2005, ApJ, 159, 197
\bibitem[]{} Falco E.E., et al., 1999, PASP, 111, 438
\bibitem[]{} Gelderman R., Whittle M., 1994, ApJS, 91, 491
\bibitem[]{} Goncalves A.C., Serote Roos M., 2004, A\&A, 413, 97
\bibitem[]{} Gu M., Chen Y., 2010, AJ, 139, 2612
\bibitem[]{} Gupta N., Srianand R., Saikia D.J., 2005, MNRAS, 361, 451
\bibitem[]{} Gupta N., Salter C.J., Saikia D.J., Ghosh T., Jeyakumar S., 2006, MNRAS, 373, 972
\bibitem[]{} Helmboldt J.F., et al., 2007, ApJ, 658, 203
\bibitem[]{} Hopkins P.F., Hernquist L., Cox T.J., Di M.T., Martini P., Robertson B., Springel V., 2005, ApJ, 630, 705
\bibitem[]{} Ishwara-Chandra C.H., Saikia D.J., 1999, MNRAS, 309, 100 
\bibitem[]{} Jamrozy M., Konar C., Machalski J., Saikia D.J., 2008, MNRAS, 385, 1286
\bibitem[]{} Junor W., Salter C.J., Saikia D.J., Mantovani F., Peck A.B., 1999, MNRAS, 308, 955
\bibitem[]{} Kim D.-C., Sanders D.B., 1998, ApJS, 119, 41
\bibitem[]{} Konar C., Jamrozy M., Saikia D.J., Machalski J., 2008, MNRAS, 383, 525
\bibitem[]{} Labiano A., et al., 2005, A\&A, 436, 493
\bibitem[]{} Labiano A., Barthel P.D., O'Dea C.P., de Vries W.H., P\'erez I., Baum, S.A., 2007, A\&A, 463, 97
\bibitem[]{} Leech K.J., Rowan-Robinson M., Lawrence A., Hughes J.D., 1994, MNRAS, 267,253
\bibitem[]{} Mantovani F., Junor W., Fanti R., Padrielli L., Saikia D.J., 1994, A\&A, 922, 59
\bibitem[]{} Mantovani F., Rossetti A., Junor W., Saikia D.J., Salter C.J., 2010, A\&A, 518, A33
\bibitem[]{} March\~a M.J.M., Browne I.W.A., Impey C.D., Smith P.S., 1996, MNRAS, 281, 425
\bibitem[]{} Mazzarella J.M., Graham J.R., Sanders D.B., Djorgovski S., 1993, ApJ, 409,170
\bibitem[]{} McMahon R.G., Irwin M.J., 1992, ASSL, 174, 417
\bibitem[]{} Moran Edward C., Halpern Jules P., Helfand David J., 1996, ApJS, 106,341
\bibitem[]{} Morganti R., Oosterloo T.A., et al., 2001, MNRAS, 323, 331
\bibitem[]{} Morganti R., Tadhunter C.N., Oosterloo T.A., 2005, A\&A, 444, 99
\bibitem[]{} Murgia M., Fanti C., Fanti R., Gregorini L., Klein U., Mack K.-H., Vigotti M., 1999, A\&A, 345, 769
\bibitem[]{} O'Dea C.P., 1998, PASP, 110, 493
\bibitem[]{} O'Dea C.P., Baum S.A., Stanghellini C., 1991, ApJ, 380, 66O
\bibitem[]{} Orienti M., Dallacasa D., Tinti S., Stanghellini C., 2006, A\&A, 450, 959
\bibitem[]{} Patnaik A.R., Browne I.W.A., Wilkinson P.N., Wrobel J.M., 1992, MNRAS, 254, 655
\bibitem[]{} Phillips R.B., Mutel R.L., 1982, A\&A, 106, 21
\bibitem[]{} Pihlstr\"om Y.M., Conway J.E., Vermeulen R.C., 2003, A\&A, 404, 871
\bibitem[]{} Polatidis A.G., Conway J.E., 2003, PASA, 20, 69
\bibitem[]{} Rees M.J., 1984, ARA\&A, 22,471
\bibitem[]{} Readhead A.C.S., Taylor G.B., Pearson T.J., Wilkinson P.N., 1996, ApJ, 460, 634
\bibitem[]{} Saikia D.J., Gupta N., 2003, A\&A, 405, 499
\bibitem[]{} Saikia D.J., Jeyakumar S., Wiita P.J., Sanghera H.S., Spencer R.E., 1995, MNRAS, 276, 1215
\bibitem[]{} Saikia D.J., Jeyakumar S., Salter C.J., Thomasson P., Spencer R.E., Mantovani F., 2001, MNRAS, 321,37
\bibitem[]{} Sanders D.B., Soifer B.T., Elias J.H., Neugebauer G., Matthews K., 1988, ApJ, 328, 35
\bibitem[]{} Snellen I.A.G., Mack K.-H., Schillizi R.T., Tschager W., 2004, MNRAS, 348, 227
\bibitem[]{} Spergel D.N., et al., 2003, ApJS, 148, 175
\bibitem[]{} Taylor G.B., Readhead A.C.S., Pearson T.J., 1996, ApJ, 463, 95
\bibitem[]{} van Gorkom J.H., Knapp G.R., Ekers R.D., Ekers D.D., Laing R.A., Polk K.S., 1989, AJ, 97, 708
\bibitem[]{} van Ojik R., R\"ottgering H.J.A., Miley G.K., Hunstead R.W., 1997, A\&A, 317, 358
\bibitem[]{} Veilleux S., Kim D.-C., Sanders D.B., 1999, ApJ, 522, 113
\bibitem[]{} Vermeulen R.C., et al., 2003, A\&A, 404, 861
\bibitem[]{} V\'eron-Cetty M.-P., V\'eron P., 2006, A\&A, 455, 773
\bibitem[]{} White R.L., Becker R.H., Helfand D.J., Gregg M.D., 1997, ApJ, 475, 479
\bibitem[]{} Wilkinson P.N., Polatidis A.G., Readhead A.C.S., Xu W., Pearson T.J., 1994, ApJ, 432, 87
\bibitem[]{} Wisotzki L., Bade N., 1997, A\&A, 320, 395
\bibitem[]{} Wright E.L., 2006, PASP, 118, 1711

\end{thebibliography}
\end{document}